# The principal astrophysical parameters of the open clusters Gulliver 18 and Gulliver 58 determined using Gaia EDR3 data

**A. L. Tadross** (altadross@nriag.sci.eg)**, E. G. Elhosseiny** (eslam_elhosseiny@nriag.sci.eg)

**National Research Institute of Astronomy and Geophysics, Cairo, Egypt.**

## Abstract

A photometric and astrometric study of the two open star clusters Gulliver 18 and Gulliver 58 was carried out for the first time using the early third data release of the Gaia space observatory (Gaia-EDR3). By studying the proper motions, parallaxes, and color-magnitude diagrams of the two clusters, we determined their actual cluster membership. Therefore, ages, color excesses, and heliocentric distances of the clusters were determined. The luminosity function, mass function, total mass, mass segregation, and relaxation time of Gulliver 18 and Gulliver 58 were estimated as well.

***Keywords***: *Galaxy: open clusters and associations; Individual: Gulliver 18, Gulliver 58; Database: GAIA; Photometry: color-magnitude diagram; Mass function; Astronomical databases: miscellaneous*

## Introduction

This article is part of our series whose goal is to use ideal contemporary datasets to obtain the basic astrophysical properties of poorly studied and/or unstudied open star clusters. Open star clusters (OCs) are the most important astronomical objects to study the Milky Way structure and evolution. OCs supply useful information about star formation mechanisms, where their main parameters, i.e., age, distance, and reddening can be derived directly from their color-magnitude diagrams (CMDs). This can be accurately achieved when we first determine the actual membership of the clusters under study (Barnes 2007; Perren et al. 2015; Bertelli et al. 2017; Marino et al. 2018).

From SIMBAD (http://simbad.cds.unistra.fr/simbad/sim-fbasic), we obtained the cluster centers in equatorial and Galactic coordinates. Gulliver 18 (hereafter G18) is located at (α = 20 h 11m 37 s, δ = +26° 31′ 55″, l = 65.526°, b = -3.97045°, J2000) in the Vulpecula constellation, whereas Gulliver 58 (hereafter G58) is located at (α = 12 h 46 m 4 s, δ = -61° 57′ 54″, l = 302.3°, b = 0.9°, J2000) in the Centaurus constellation. **Fig. 1** shows the negative images of G18 and G58 as taken from ALADIN at DSS-colored optical wavelengths. Cantat-Gaudin et al. (2018) studied the main astrometric parameters of those two clusters, i.e., coordinates, proper motions, parallaxes, and distances as newly discovered clusters in the Milky-Way Galaxy. They used the second data release of the Gaia DR2 database, Gaia Collaboration et al. (2018).

Here, we estimated the fundamental parameters of the two clusters G18 and G58 for the first time using the early third release of the Gaia database (Gaia EDR3) - Gaia Collaboration et al. (2021), which was published on December 3, 2020. Gaia EDR3 offers improved astrometry and photometry for 1.8 billion sources brighter than G ≈ 21 mag. Compared to Gaia DR2, the parallax enhancement is 20% and the proper motions are twice more accurate. The most important part of using the Gaia EDR3 lies in five astrometric parameters: equatorial positions (α, δ), proper motions ($\mu_\alpha \cos\delta$, $\mu_\delta$), and parallaxes (ϖ). In addition, the magnitudes in three photometric filters (G, $G_{BP}$, $G_{RP}$) were obtained with better homogeneity due to the significant advance in several aspects (Gaia Collaboration et al. 2021; Torra et al. 2021; Riello et al. 2021). Of course, all of these improvements affect the astrophysical estimated parameters of the clusters under study.

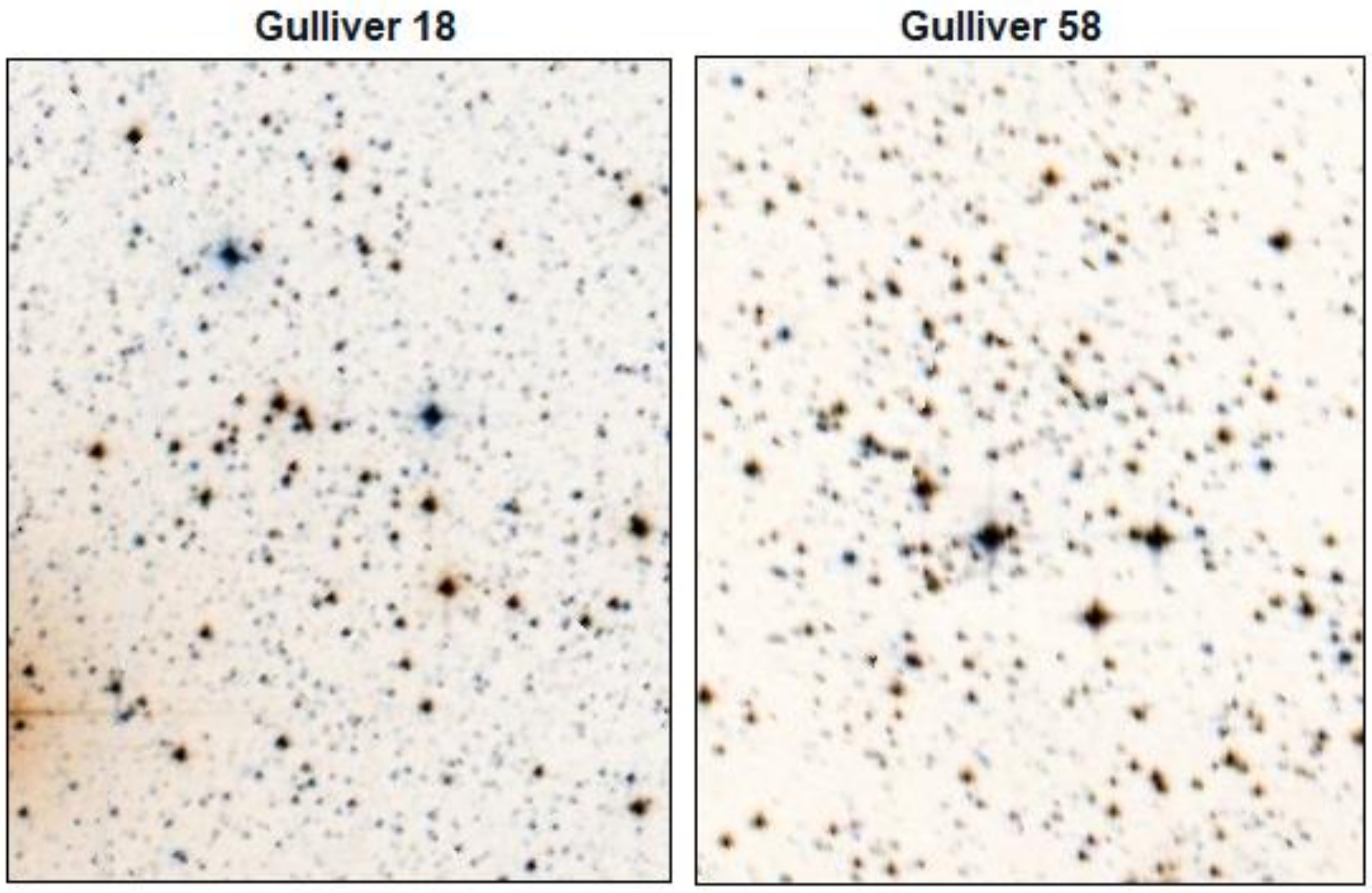

Fig. 1. Inverse colored (negative) images of the clusters G18 and G58 as taken from ALADIN at DSS colored optical wavelength. These two clusters lie in the Vulpecula and Centaurus constellations respectively. North is up and East to the left.

The paper is organized as follows: the Gaia EDR3 dataset and membership determination are presented in Section 2. Section 3 shows the angular size according to the radial density profiles (RDPs). Section 4 contains the photometry of the color-magnitude diagrams (CMDs). Section 5 describes the clusters' dynamic states, i.e., luminosity functions (LFs), mass functions (MFs), mass segregation and relaxation times. Section 6 summarizes the findings results and conclusions.

## 2. Gaia EDR3 Dataset and membership

The standard dataset of G18 and G58 was downloaded from the Gaia EDR3 I/350 Vizier catalog website. A circular region of 20 arcmin radius centered in the celestial position was applied to each cluster. The error ranges of the parallaxes are up to 0.03 mas for G<15 mag, 0.07 mas for G≈17 mag, 0.5 mas for G≈20 mag, and 1.3 mas for G≈21 mag. The error ranges of the proper motions (PMs) are up to 0.03 mas/yr for G<15 mag, 0.07 mas/yr for G≈17 mag, 0.5 mas/yr for G≈20 mag, and 1.4 mas/yr for G≈21 mag.

Using high-precision Gaia EDR3 parallaxes and proper motions, we can easily remove the background field stars from the cluster's main sequence (Bellini et al. 2009; Yadav et al. 2013; Sariya and Yadav 2015; Tadross 2018). The Vector Point Diagrams (VPDs), $\mu\alpha \cos \delta$ vs. $\mu\delta$, of G18 and G58 are shown in Fig. 2. The greatest density area (the darkest spot) is taken as the subset of the cluster's most likely members (Tadross 2018; Tadross and Hendy 2021 & 2022).

In our analysis, we used the software named ***TopCat***. This is a tool that can be handling huge and sparse datasets. It was initially created for astronomy to support virtual observatories. The acronym ***TOPCAT*** derives from **T**ool for **OP**erations on **C**atalogues **A**nd **T**ables. It can support several digital file formats, including FITS, which is widely used in astronomy (http://www.star.bris.ac.uk/~mbt/topcat/).

Within the subset of the cluster's most likely members, only those stars with magnitudes G ≤ 20.5, located inside the cluster's estimated size (see Section 3) were taken into account. Mean values and standard deviations of the parallax and the two components of the proper motions were calculated for all these stars (excluding negative values of the parallax, see the lower right-hand panel of **Fig. 4**). Stars are then considered to be cluster members only if their 3σ parallax and proper motion errors lie within the cluster's mean values with respect to the background field ones. According to Lindegren et al. (2018), all the parallax values should be shifted by adding 0.029 mas to their values. In addition, the value of the **R**enormalized **U**nit **W**eight **E**rror, **RUWE**, indicates how well the source matches the single-star model - it should be less than 1.4.

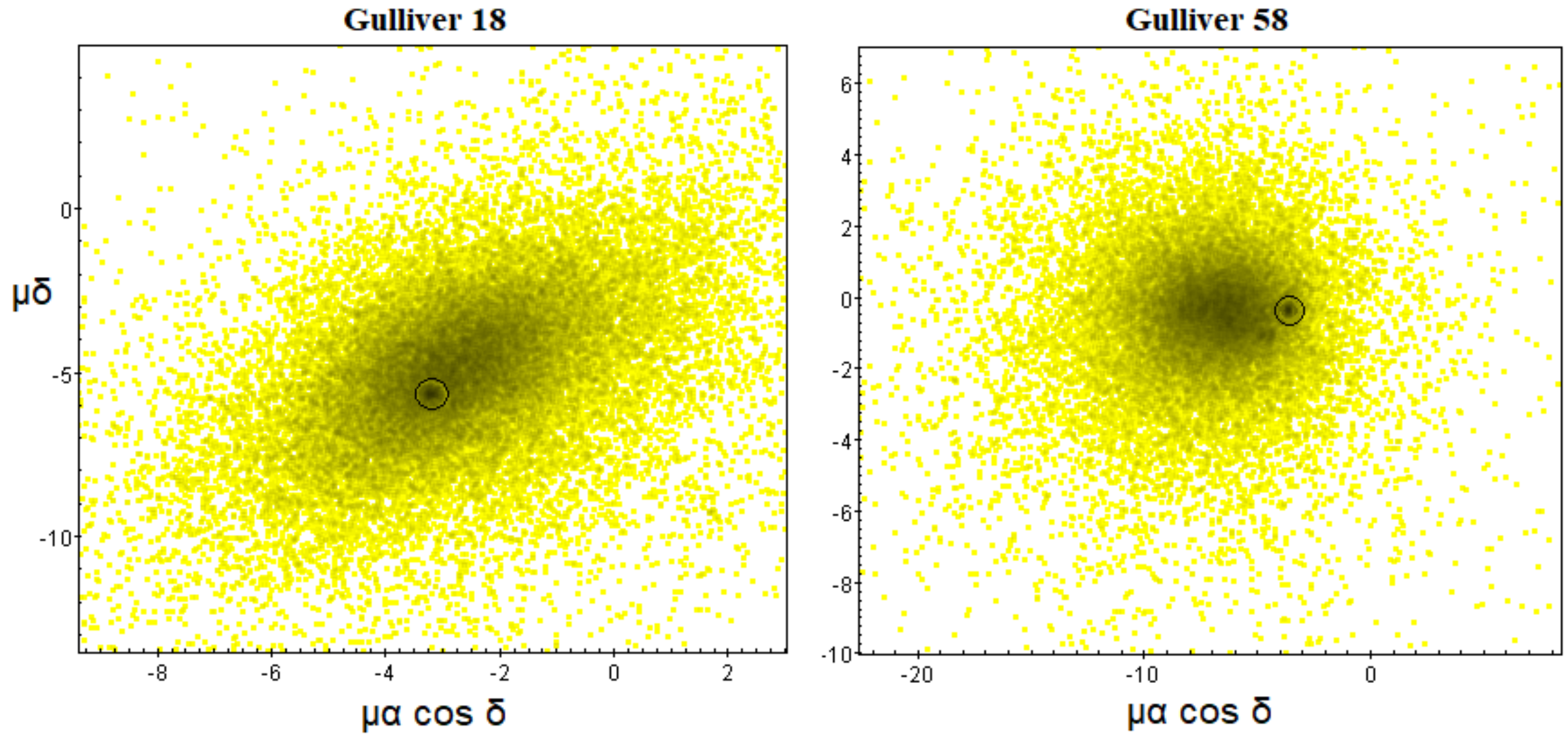

Fig. 2. The vector point diagrams of the clusters G18 and G58. The circles denote the darkest areas, in which the subsets of the most likely members lie.

Using TopCat, we can identify comoving stars, i.e., those stars that move at the same speed and direction in the sky, as shown in **Fig. 3**. It's worth noting that the selected subset of VPDs influences the CMDs and field star separation. The CMDs of the clusters appear cleaner when those conditions are applied (Anderson et al. 2006; Sariya et al. 2017).

## 3. Angular Size

The boundary and core radii of G18 and G58 were calculated using King's (1966) radial density profile (RDP). To do so, we built a series of concentric circles centered on the clusters' central coordinates. The number of stars found in each ring was divided by the ring area to obtain the stellar density *f(R)*. The upper right-hand panel of **Fig. 4** shows the measured star density vs. the distance to the cluster center. The King model fit can be given by the equation:

$$f(R) = f_{bg} + \frac{f_0}{1 + \left(\frac{R}{R_c}\right)^2}$$

where R is the radius from the cluster center, $f_0$ the central density, $R_c$ the core radius and $f_{bg}$ the background density. We can find these parameters from the King model as follows:

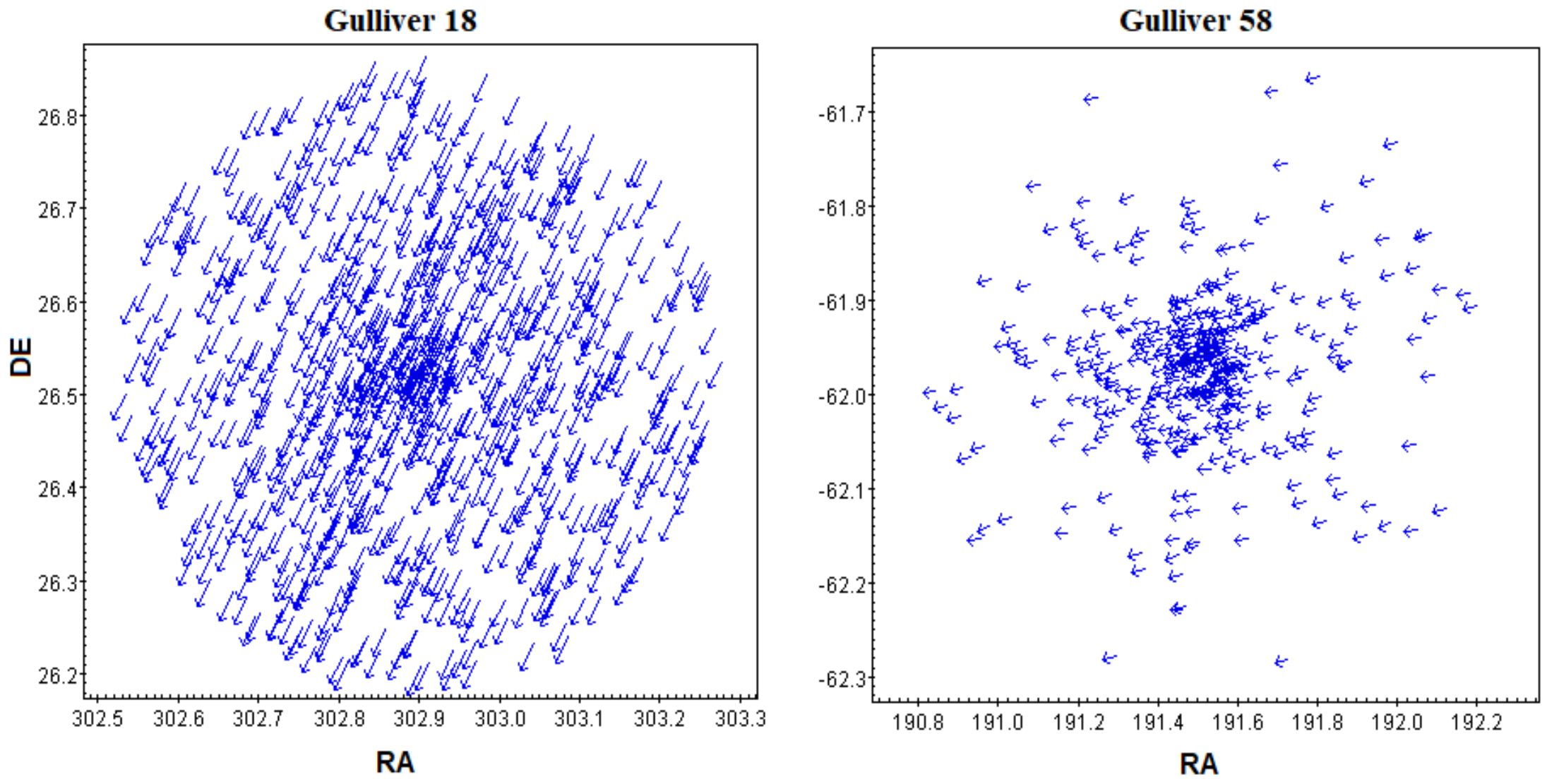

Fig. 3. Comoving stars of clusters G18 and G58. They are the stars included in the subsets we selected in the VPDs, i.e., the stars that move at the same speed and direction in the sky.

| | $R_c$ | $f_{bg}$ | $f_0$ |
|---|---|---|---|
| **G18** | 0.35 | 50 | 233 |
| **G58** | 0.27 | 33 | 305 |

When the stars approach the boundary radius, they begin to dissolve within the background density. Knowing the previous parameters, we can get the boundary radius by applying the equation of Bukowiecki et al. (2011) as follows:

$$R = R_c \sqrt{\frac{f_0}{3\sigma_{\rm bg}} - 1}$$

where $\sigma_{\rm bg}$ is the uncertainty of $f_{bg}$. Consequently, the estimated boundary radii of G18 and G58 are found to be 7.5 ± 0.5 arcmin (3.0 pc) and 4.5 ± 0.3 arcmin (1.9 pc), respectively.

On the other hand, the tidal radius of a cluster is defined as the distance from the cluster core at which the Galaxy's gravitational influence equals that of the cluster core. Estimating the total masses of G18 and G58 (see Sec. 5), the tidal radius can be calculated using the form of Jeffries et al. (2001) as follows:

$$R_t = 1.46 \times M_c^{1/3}$$

where $R_t$ is the tidal radius (in parsec) and $M_c$ the total mass of the cluster (in solar masses). The tidal radii of G18 and G58 are found to be 17.8 and 11.0 (± 0.2) pc, respectively. The concentration parameter of Peterson & King (1975), $C = \log\left(\frac{R}{R_c}\right)$, shows us how the cluster is prominent or condensed with respect to the background field stars. We found that Both clusters are insufficiently condensed objects.

**4. CMD photometry**

The study of CMD is a widely used technique to the observed main sequence of the cluster. This is achieved by finding the best fit of one of the Padova PARSEC databases of stellar evolutionary isochrones (http://stev.oapd.inaf.it/cgi-bin/cmd) of Bressan et al. (2012) to the observed main-sequence curve of the cluster. To decrease the effect of field stars contamination, only the most likely cluster members are taken into account. Since the cluster members have a common origin, they share the same speed and direction in the sky. This fact makes proper motions a valuable tool for removing nonmember stars from each cluster's main sequence (Yadav et al. 2013; Bisht et al. 2020).

Based on the CMDs of G18 and G58, as shown in the left-hand panel of **Fig. 4**, only the stars within the cluster's boundary size and parallaxes ranges are represented. The mean values of the parallaxes are found to be 0.50 and 0.45 (± 0.05) mas, respectively. Clusters fittings were quite good by shifting the isochrones of ages 100 ± 10 Myr and 1.0 ± 0.1 Gyr with apparent distance moduli (m-M) of 12.3 and 13.6 (± 0.25) mag, respectively. In addition, the color excesses *E(BP-RP)* are found to be 0.8 and 1.4 (± 0.10) mag, respectively. Note that G18 and G58 are close to areas with a lot of dust with high differential extinction in their fields along their lines of sight, as displayed in the 3D extinction map (http://argonaut.skymaps.info).

Reddening is a critical parameter affecting the total absorption value that must be subtracted from the apparent distance modulus to obtain the true distance to the cluster. We used the Padova PARSEC database of stellar evolutionary tracks and isochrones, which is scaled to the solar metallicity of 0.0152. The Gaia filter passbands are taken from Riello et al. (2021), where $A_G/A_v$= 0.836, $A_{GBP}/A_v$=1.083, and $A_{GRP}/A_v$=0.634. These ratios have been used for correction of the magnitudes for the interstellar reddening and converting the color excess to E(B-V), where $R_v$ = $A_v$/E(B-V)=3.1. Therefore, we can estimate the true distance moduli $(m\text{-}M)_0$ of G18 and G58, from which we infer heliocentric distances of 1370 and 1425 (± 65) pc, respectively. Correspondingly, the Cartesian Galactocentric coordinates ($X_\odot$; $Y_\odot$; $Z_\odot$) and the distances from the Galactic center ($R_g$) are estimated for the two clusters as listed in **Table 1.** According to our calculations mentioned in Tadross (2011), the Y-axis connects the Sun to the Galactic center, being positive to the Galactic anticenter, while the X-axis is perpendicular to Y-axis, being positive in the first and second Galactic quadrants (Lynga 1982). We adopted a Galactocentric distance ($R_g$)of 7.2 kpc (Bica et al. (2006).

## 5. The Dynamic State of the Clusters

### 5. 1. Luminosity and mass functions

We used the photometric dataset of Gaia EDR3 to derive the clusters' luminosity functions (LFs) and mass functions (MFs). The LF represents the distribution of the absolute magnitudes of the cluster's members. Using the distance moduli obtained from the isochrone fittings, we transformed the apparent G magnitudes of the cluster members into absolute magnitudes. Then, the LF diagrams can be constructed as shown in the upper panel of **Fig. 5**. Note that the LFs increase up to $M_G \sim 8.15$ and 4.20 mag for G18 and G58, respectively.

The initial mass function (IMF) provides the main bond between the bright massive members and less massive fainter ones. It is a historic record of the star formation process and plays the main role in understanding the early dynamic development of star clusters. IMF was estimated for the bright massive stars ($\geq 1\ M_{\odot}$) by Salpeter's (1955) power law, where the number of stars in each mass range decreases as the mass increases. It can be written as follows:

$$\boldsymbol{log\frac{dN}{dM} = -(1+x)\log(M) + const.}$$

where $\boldsymbol{dN}$ is the number of stars in a mass bin $\boldsymbol{dM}$ with a central mass $\boldsymbol{M}$ and $\boldsymbol{x}$ is the MF slope. To convert LF into MF, we used the last version of the theoretical isochrones of Padova's stellar evolutionary tracks and isochrones. The resulting mass functions of G18 and G 58 are shown in the lower panel of **Fig. 5**. The derived values of the MF slope are found to be $x$ = 2.25 and 2.29 (± 0.15) for G18 and G58, respectively, which agree with Salpeter's mean value.

### 5. 2. Mass segregation

Mass segregation in a real cluster implies that the massive stars are much more integrated toward the cluster center than the lower mass stars. Mass segregation is a result of the dynamic evolution of the cluster and/or an impression of star construction processes themselves or both, Sagar (2002). To explore if there is actual mass segregation, we divided the clusters' stars into four bands, G<17; 17≤G≤18; 18≤G≤19 and G≥19 mag. Drawing these bands as a function of the distances from the cluster's center as shown in **Fig. 6,** we found that the bright massive stars are more likely to settle toward the cluster center than the less massive fainter ones.

### 5.3. Relaxation time

Once the distribution of the cluster members' velocities becomes almost Maxwellian, a metric for understanding the dynamical evolution is considered. This period is known as "Relaxation Time" ($\boldsymbol{T_R}$) and can be defined by the equation of Spitzer and Hart (1971) as follows:

$$\boldsymbol{T_R = \frac{8.9 \times 10^5 \sqrt{N} \times R_h^{1.5}}{\sqrt{\langle m \rangle} \times \log(0.4N)}}$$

where $\boldsymbol{N}$ is the number of cluster members, $\boldsymbol{R_h}$ is the cluster's radius that contains half of the cluster's total mass (in pc) and $\boldsymbol{\langle m \rangle}$ is the average mass of a member star (in solar masses). Thus, the dynamic relaxation times are found to be 31.5 and 9.5 (± 5.0) Myr for G18 and G58, respectively. The clusters under investigation are thus found to be older than their estimated relaxation times. We conclude that G18 and G58 are dynamically relaxed clusters.

## 6. Results and conclusions

The principal properties of the two open clusters Gulliver 18 and Gulliver 58 were looked into in this paper for the first time using the Gaia EDR3 database. Both clusters are located in the Vulpecula and Centaurus constellations, respectively, while they are not sufficiently condensed objects in the sky. The MF slopes of G18 and G58 are found to agree with Salpeter's (1955) mean value. The two clusters are dynamically relaxed as their estimated relaxation times are much smaller than their ages. **Table 1** summarizes the main results of our study.

**Table 1. The astrophysical principal parameters of G18 and G58.**

| Parameter | Gulliver 18 | Gulliver 58 |
|---|---|---|
| **RA** *(h: m: s)* | 20:11:37 | 12:46:04 |
| **DE** *(º: ’: ”)* | +26:31:55 | -61:57:54 |
| **G. long.** *(º)* | 65.526 | 302.30 |
| **G. lat.** *(º)* | -3.97 | 0.90 |
| **Age** *(Myr)* | 100 ± 10 | 1000 ± 100 |
| **Radius** *(arcmin)* | 7.5 ± 0.5 | 4.5 ± 0.3 |
| **Core Radius** *(arcmin)* | 0.35 ± 0.07 | 0.27 ± 0.04 |
| **Tidal Radius** *(pc)* | 17.8 ± 0.5 | 11.0 ± 0.5 |
| **m-M** *(mag)* | 12.3 ± 0.25 | 13.6 ± 0.25 |
| **E(BP-RP)** *(mag)* | 0.8 ± 0.1 | 1.4 ± 0.1 |
| **E(B-V)** *(mag)* | 0.61 ± 0.1 | 1.06 ± 0.1 |
| **Dist.** *(pc)* | 1370 ± 65 *(1558.6)* | 1425 ± 65 *(2344.2)* |
| **Relax. Time** *(Myr)* | 31.5 ± 5 | 9.5 ± 5 |
| **P.M.** *(mas/sec)* | 6.515 ± 0.45 *(6.488)* | 3.652 ± 0.35 *(3.609)* |
| **Plx.** *(mas)* | 0.50 ± 0.07 *(0.613)* | 0.45 ± 0.07 *(0.398)* |
| **Rgc** *(kpc)* | 6.75 ± 0.2 | 6.55± 0.2 |
| $X_\odot$ *(pc)* | -565 ± 40 | -760 ± 60 |
| $Y_\odot$ *(pc)* | 1241 ± 55 | -1201 ± 50 |
| $Z_\odot$ *(pc)* | -95 ± 5 | 22 ± 5 |

The values in brackets are the corresponding astrometric measurements obtained by Cantat-Gaudin et al. (2018).

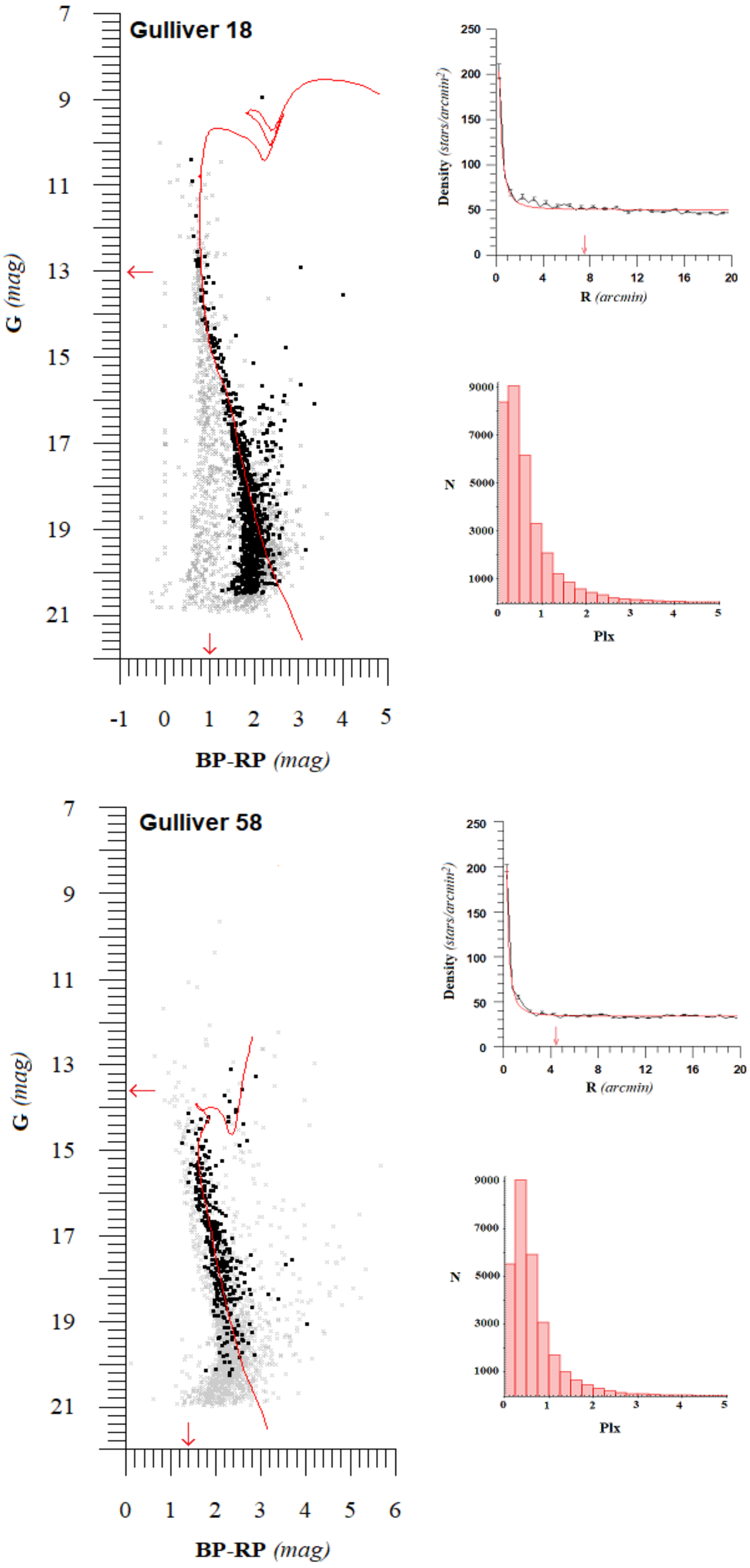

Fig. 4. The left-hand panels show the main sequence curves of G18 and G58 fitted to the solar metallicity theoretical isochrones of Padova. The ages are found to be 100 ± 10 Myr and 1.0 ± 0.1 Gyr, respectively. The distance moduli and color excesses are found to be 12.3, 13.6 (± 0.25) mag, 0.8 and 1.4 (± 0.10) mag, respectively. The right-hand panels show the radial density profiles fitted to the King (1966) model and the histograms of parallaxes of the selected stars (ϖ>0). The estimated radii of G18 and G58 are found to be 7.5 and 4.5 (± 0.50) arcmin and the mean values of the parallaxes are found to be 0.50 and 0.45 (± 0.05) mas, respectively.

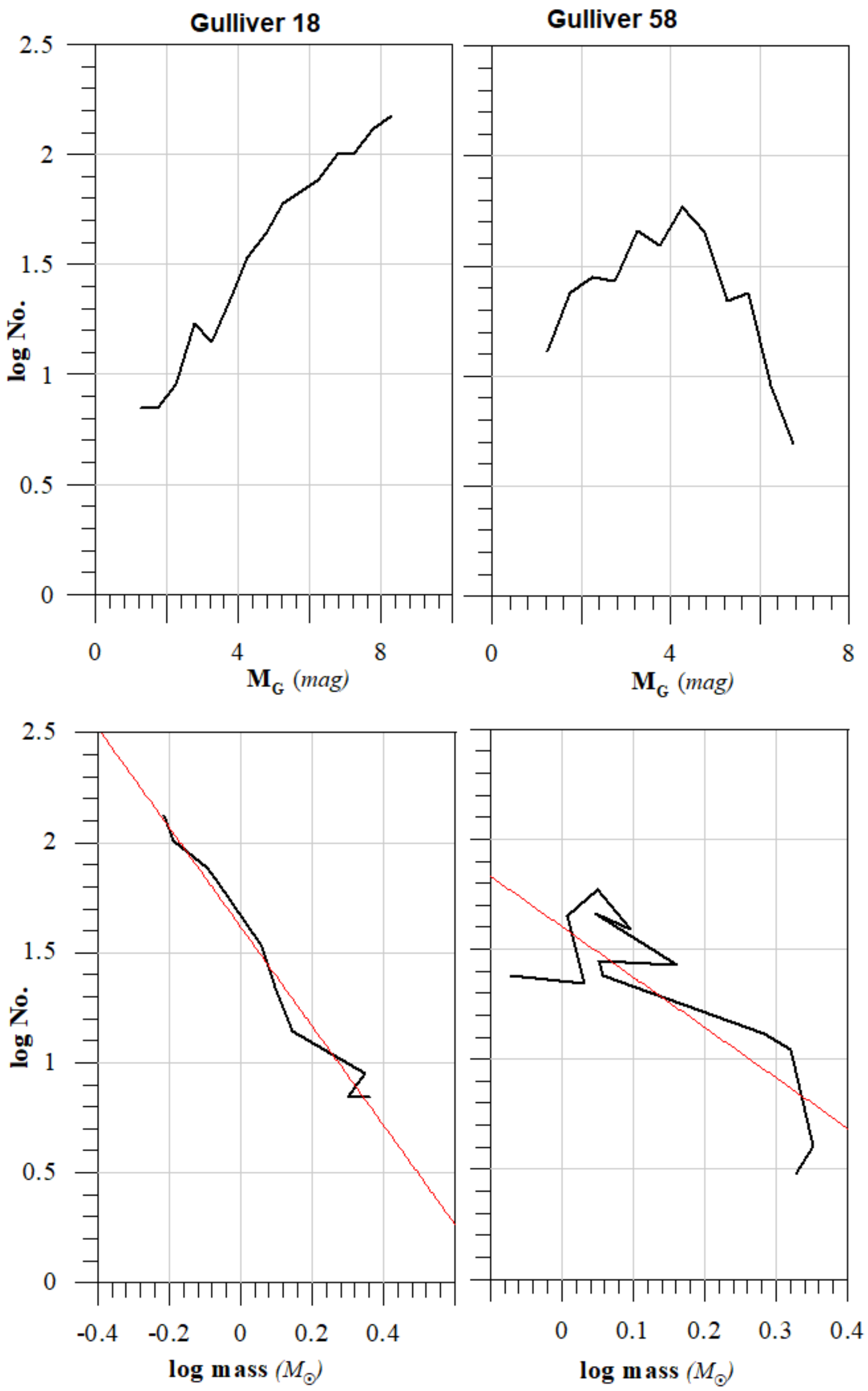

Fig. 5. The upper panel represents the luminosity functions of G18 and G58. Note that the LFs increase up to $M_G \sim 8.15$ and 4.20 mag for G18 and G58, respectively. The lower panel represents the mass distribution of the two clusters. The red lines show the linear fittings of the mass functions' slopes, which are found to be 2.25 and 2.29 (± 0.15) for G18 and G58, respectively.

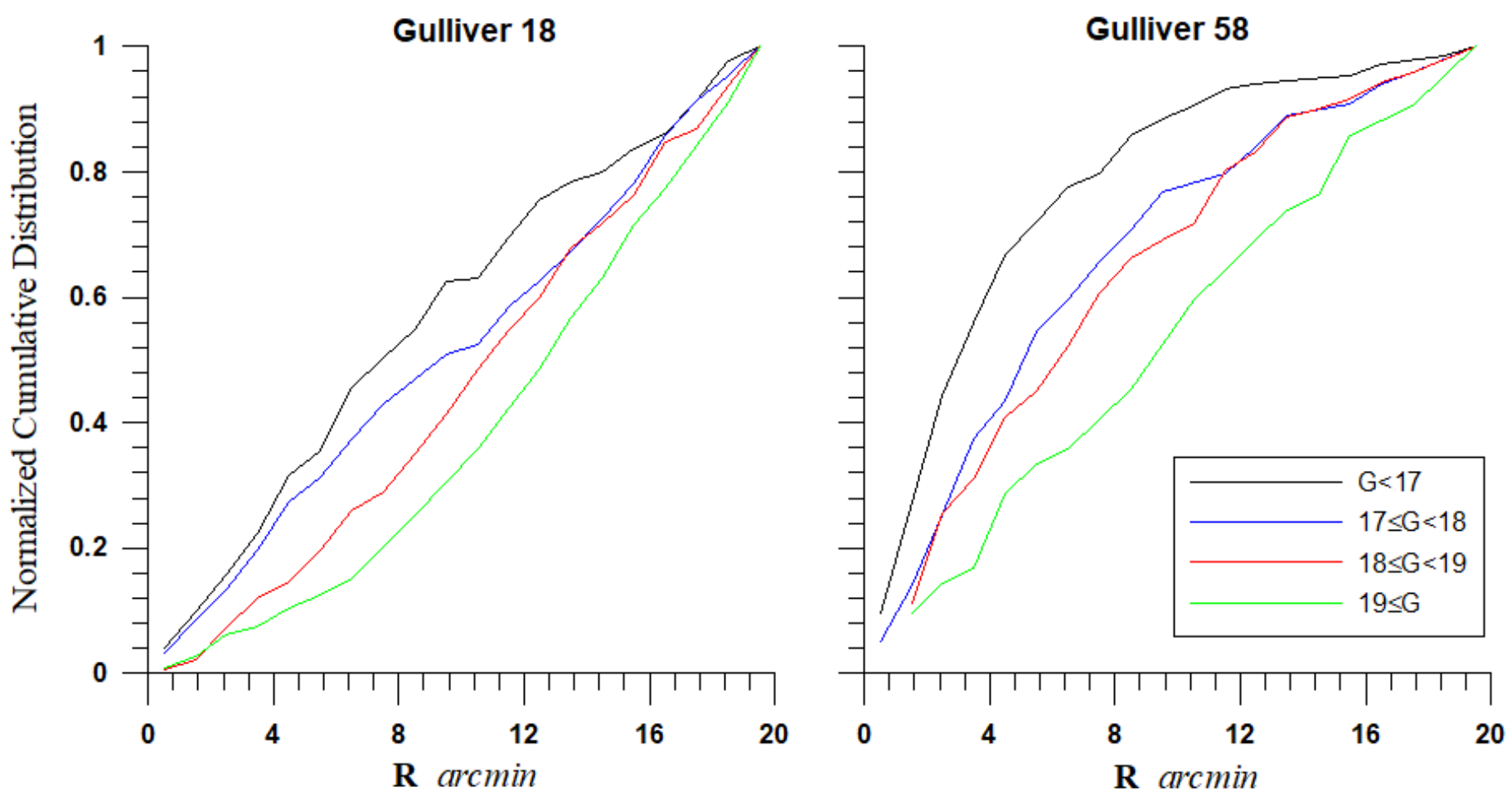

Fig. 6: The accumulated frequency distributions of G18 and G58 represent the radial distance and magnitude distributions of the clusters' member bands (mass segregation). The bright, massive stars are more likely to settle toward the clusters' centers than the fainter less massive ones.

**Acknowledgments**

This work is a part of the project named ''IMHOTEP'' No. 42088ZK between Egypt and France. It was begun in 2019 and finished in 2021. Many thanks to Prof. David Valls-Gabaud (Paris Observatory) for being a companion in that project. This work has made use of data from the European Space Agency (ESA) mission Gaia processed by the Gaia Data Processing and Analysis Consortium (DPAC), (https://www.cosmos.esa.int/web/gaia/dpac/consortium). Funding for the DPAC has been provided by national institutions, in particular, the institutions participating in the Gaia Multilateral Agreement (MLA).
The Gaia mission website is https://www.cosmos.esa.int/gaia
The Gaia archive website is https://archives.esac.esa.int/gaia